\documentclass[fleqn,usenatbib]{mnras}
\usepackage{newtxtext,newtxmath}
\usepackage[T1]{fontenc}
\DeclareRobustCommand{\VAN}[3]{#2}
\let\VANthebibliography\thebibliography
\def\thebibliography{\DeclareRobustCommand{\VAN}[3]{##3}\VANthebibliography}

\usepackage{graphicx}	% Including figure files
\usepackage{amsmath}	% Advanced maths commands
\usepackage{subcaption}
\usepackage{cleveref}
\AtBeginDocument{\mathcode`v=\varv}
\usepackage{comment}
\usepackage{adjustbox}

\title[Simulating the Stellar Bycatch within Radio SETI Surveys]{Simulating the Stellar Bycatch: Constraining the Prevalence of Extraterrestrial Transmitters within Radio SETI Surveys}
\author[L. Mason et al.]{Louisa A. Mason$^{1}$, Michael A. Garrett,$^{1,2,6}$ and Andrew P. V. Siemion $^{1,3,4,5,6}$
\\
$^{1}$Jodrell Bank Centre for Astrophysics, Department of Physics and Astronomy, Alan Turing Building, University of Manchester, Oxford Road, M13 9PL, UK\\
$^{2}$Leiden Observatory, Leiden University, P.O. Box 9513, NL-2300 RA Leiden, The Netherlands\\
$^{3}$Astrophysics Sub-Department, Department of Physics, University of Oxford, Denys Wilkinson Building, Keble Road, Oxford OX1 3RH, UK\\
$^{4}$SETI Institute, 339 Bernardo Avenue, Suite 200, Mountain View, CA 94043, USA\\
$^{5}$Berkeley SETI Research Center, University of California, Berkeley, CA 94720, USA\\
$^{6}$University of Malta, Institute of Space Sciences and Astronomy, Msida, MSD2080, Malta\\
}

\date{Accepted 2025 November 21. Received 2025 November 7; in original form 2025 August 20}

\pubyear{2025}

\begin{document}

\label{firstpage}
\pagerange{\pageref{firstpage}--\pageref{lastpage}}
\maketitle

\begin{abstract}
Searches for radio technosignatures place constraints on the prevalence of extraterrestrial transmitters in our Galaxy and beyond. It is important to account for the complete stellar population captured within a radio telescope's field of view, or stellar 'bycatch'. In recent years, catalogues from ESA’s Gaia mission have enabled SETI surveys to place tighter limits on extraterrestrial transmitter statistics. However, Gaia remains 
restricted by magnitude limits, astrometric uncertainty at large distances, and confusion in crowded regions. To address these limitations, we investigate the use of the Besançon Galactic Model to simulate the statistical underlying stellar population to derive more realistic constraints on the occurrence of extraterrestrial transmitters. We apply this method to Breakthrough Listen's Enriquez/Price survey, modelling 6,182,364 stellar objects within 1229 individual pointings and extending the search out to distances $\leq 25$~kpc. We place limits on the prevalence of high duty cycle transmitters within 2.5~kpc, suggesting  $\leq (0.000995 \pm 0.000002) \%$ of stellar systems contain such a transmitter (for near-zero drift rates and EIRP$_{\mathrm{min}} \gtrsim 5 \times 10^{16}$~W). In support of broader adoption, we provide a simple calculator tool that enables other researchers to incorporate this approach into their own SETI analyses. Our results enable a more complete statistical estimation of the number and stellar type of systems probed, thereby strengthening constraints on technosignature prevalence and guiding the analysis of future SETI efforts. We also conclude that SETI surveys are, in fact, much less biased by anthropocentric assumptions than is often suggested.

\end{abstract}


%%%%%%%%%%%%%%%%%%%%%%%%%%%%%%%%%%%%%%%%%%%%%%%%%%

%%%%%%%%%%%%%%%%% BODY OF PAPER %%%%%%%%%%%%%%%%%%

\section{Introduction}

The search for extraterrestrial intelligence (SETI) — whether for intentional beacons, accidental leakage, or relic signals from extinct civilisations — spans a vast and complex parameter space. This includes dimensions such as position on the sky, observing frequency, total bandwidth, spectral resolution, temporal cadence, signal modulation, and sensitivity, as discussed by \citet{wright_how_2018}. SETI surveys aim to reduce the volume of unsearched multidimensional parameter space to tighten the bounds on the prevalence of extraterrestrial transmitters. 

Narrowband single-dish SETI surveys typically search for signals across a broad range of Doppler drift rates (e.g. $\pm 4$~Hz/s) and possess a field of view at 1~GHz that spans several arcminutes. %, even for the largest radio telescopes. 
Each pointing thus encompasses a significant number of background (and occasionally foreground) celestial objects ($N_*$) which are passively included in the analysis of the integrated power spectrum. As a result, these surveys are sensitive not only to signals originating from the nominal target star, but also to potential transmitters distributed across the full field of view. Quantifying this ‘stellar bycatch’ — the incidental population of stars within the telescope beam — enables more realistic constraints to be placed on the prevalence of extraterrestrial transmitters probed by SETI surveys (see \citet{wlodarczyk-sroka_extending_2020}).

ESA’s \textit{Gaia} mission and its associated astrometric catalogue have enabled better estimates of the Galactic stellar bycatch for the original Breakthrough Listen surveys by \citet{enriquez_breakthrough_2017} and \citet{price_breakthrough_2020} (see \citet{wlodarczyk-sroka_extending_2020}). This methodology has since been adopted by other studies, including those by \citet{margot_search_2023}, \citet{johnson_simultaneous_2023}, and \citet{mason_conducting_2025}. The inclusion of the stellar bycatch increases the total population of stars surveyed within a given pointing direction and observational time frame. 

However, the latest \textit{Gaia} data release includes only 1.8 billion Galactic objects \citep{gaia_collaboration_gaia_2023} and remains limited by magnitude limits, astrometric uncertainties, and confusion in densely populated regions. Catalogues derived from \textit{Gaia}, such as those by \citet{bailer-jones_estimating_2021} represent a lower bound on the true stellar bycatch encountered in SETI surveys.

By utilising galactic stellar population synthesis models, we can not only predict the number distribution of stars as a function of distance for a particular cut through the Milky Way, but also yield the range of stellar classes that a typical survey passively includes. Even strictly “targeted” single-dish SETI surveys nominally focused on specific stellar type (such as nearby Main Sequence stars) invariably encompass a much broader array of spectral types through their incidental stellar bycatch. Without accounting for this broader context, we risk underestimating the true extent and scientific reach of SETI survey efforts.

In this paper, we use the Besançon Galactic Model (BGM) to simulate stars within the field of view of SETI surveys in order to place better limits on the prevalence of extraterrestrial transmitters in the Milky Way. In section 2, we discuss the limitations of using \textit{Gaia} alone to estimate the stellar bycatch population and briefly outline the BGM's simulation method. In section 3, we utilise the BGM to simulate the stellar population associated with the \citet{price_breakthrough_2020} survey and compare the limits on prevalence and stellar population properties calculated against previous studies utilising \textit{Gaia} by \citet{wlodarczyk-sroka_extending_2020}. In section 4, we provide a public domain calculator that estimates the stellar bycatch as a function of pointing position and field of view. In section 5, we present our conclusions and evaluate the significance of employing simulations to estimate the stellar bycatch for future SETI surveys. 

\section{Modelling the Milky Way}

\subsection{Limitations of Gaia}
ESA’s \textit{Gaia} mission has transformed our understanding of the Milky Way by providing astrometric measurements for over a billion stellar objects.  However, like all observational datasets, its completeness and reliability are subject to important limitations.

The mission achieves high completeness for stars with apparent magnitudes in the range $12 < G < 17$, and can extend to $G \sim 20$ in regions of low stellar density. In contrast, heavily crowded fields with more than 400,000 stars per square degree see completeness drop to $G \sim 18$ \citep{luri_gaia_2018}. Distance estimates are also affected by parallax uncertainties; for sources with a fractional parallax error $f > 0.2$, geometric distance estimates such as those provided by \citet{bailer-jones_estimating_2021} are preferred. In the cases of large or negative fractional parallax errors, % ($f < 0$ or $f > 1$),
inferred distances are dominated by modelling prior errors. This limits reliable distance estimation within the \textit{Gaia} catalogue (where $0 < f < 1$) to within 10~kpc.

The classification of stellar types in \textit{Gaia} is typically inferred from photometric colour, using the difference $G_{BP} - G_{RP}$. However, in crowded regions or for faint sources, background contamination and confusion — particularly from unresolved binaries — can significantly degrade the accuracy of spectral classification \citep{arenou_gaia_2018}. Distinguishing between nearby Main Sequence stars and White Dwarfs, for instance, is often problematic in these regimes.

To mitigate these issues, stringent quality filters are applied to extract clean stellar samples. Yet this comes at a significant cost: only about 30\% of objects in \textit{Gaia} DR2 pass the recommended contamination filters \citep{arenou_gaia_2018}. As a result, while \textit{Gaia} provides an invaluable foundation for stellar population studies, it offers only a lower limit on the true number of stars intercepted by wide-field SETI observations.

\subsection{The Besançon Galactic Model}

The BGM (\citet{robin_synthetic_2004}, \citet{bienayme_new_2018}) is a dynamical self-consistent model of the Milky Way. The model is used across astrophysics research, including studies of Galactic structure, stellar populations (in particular for low mass stars), extinction mapping and microlensing predictions (\textit{e.g.} \citet{reyle_milky_2009}; \citet{czekaj_besancgalaxy_2014}; \citet{ravinet_study_2022} and \citet{specht_mabls-2_2020}). In the context of stellar population synthesis, the BGM has proven especially valuable for modelling star counts, kinematics, and photometric distributions across a range of sky directions. 

The model divides the Galaxy into 4 subcomponents: the thin disc, the thick disc, the bulge and stellar halo. The thin disc has been divided into 7 subpopulations based on age, each modelled by an Einasto relation describing the 3D density distribution \citep{einasto_galactic_1979}. The thick disc is modelled by two isochrones, due to the 2GYr long star formation period. The halo can be considered a homogeneous population of stars with a short star-forming period. A Salpeter IMF is assumed for the bulge, despite the difficulty in constraining low mass stars within the bulge (due to crowding).

The stellar densities are constrained by the galactic potential (poisson equation), velocity dispersion of each subcomponent (based on the age-velocity dispersion relation) and the galactic rotation curve (to estimate the scale heights/axis ratio of each subpopulation). The model is initially based on the stellar neighbourhood, before scaling outwards to the full extent of the Milky Way.

The population of stars within each subcomponent of the galaxy can be estimated given, 
\begin{equation}
    N_* = \int\Phi(m) dm \int \Psi(t) dt
\end{equation}
where $\Phi(m)$ is the initial mass function and $\Psi(t)$ is the star formation rate \citep{haywood_model_1994}. For a number of stars per age and mass interval, the BGM uses evolutionary tracks to evaluate further stellar parameters. If interpolating along the evolutionary tracks results in a solution, the star's luminosity $L$, surface gravity $g$ and effective temperature $T_{\mathrm{eff}}$ are assigned; if not, the star is considered a 'stellar remnant'.
%If a star's evolutionary track finishes on the HR diagram, the star's luminosity $L$, surface gravity $g$ and effective temperature $T_{\mathrm{eff}}$ are assigned; if the track doesn't finish within the HR diagram, the star is considered a 'stellar remnant'. 
For stars with assigned $L$, $g$ and $T_{\mathrm{eff}}$, the photometric colour is estimated using an atmosphere model (BaSeL). Reddening is considered using 3D extinction models to estimate the apparent magnitude of stars based on the line of sight through the galaxy. 

The BGM has been constrained against several surveys: \textit{Hipparcos/ Tycho} for the local vicinity/self-consistent densities, \textit{Gaia}, \textit{RAVE} and \textit{2MASS} for the outer disc scales (see \citet{robin_self-consistent_2022}, \citet{czekaj_besancgalaxy_2014} \& \citet{deforet_3d_2024} for a sample of the extensive updates to this galactic synthesis model from observational datasets). White dwarfs are modelled from cooling tracks, and classified into three broad populations: Hydrogen rich white dwarfs (labelled DA), helium rich white dwarfs (DB) and continuous spectrum white dwarfs (DC). AGB stars are added empirically but brown dwarfs are not modelled by the BGM.

The BGM models the stellar population down to the hydrogen burning limit of red dwarfs (M > 0.12~M.). However, the assumed IMFs are poorly constrained for low mass stars and likely contaminated by galaxies for V > 22 \citep{robin_synthetic_2004}. Improved 3D extinction maps remains limited for lines of sight of high density and insufficient completeness, such as within the Galactic Centre and spiral arm tangents \citep{marshall_modelling_2006}. 

\section{Application to Current SETI Surveys}

\subsection{The Enriquez/Price Survey} \label{sec:application}

\citet{enriquez_breakthrough_2017} conducted a survey of 692 stars using the 100-m Robert C. Byrd Green Bank Telescope (GBT) at L-band. The target sample was derived from the list proposed by \citet{isaacson_breakthrough_2017} for the Breakthrough Listen initiative, which included the 60 closest stars, a range of \textit{Hipparcos} stars across various spectral classes, and a selection of exotic objects (\textit{e.g.} white dwarfs, neutron stars, asteroids). \citet{price_breakthrough_2020} extended this survey by observing 1327 stars from Isaacson's list, again using the GBT at L-band (1.10 - 1.90~GHz) and S-band (1.80 - 2.80 GHz), as well as the 64-m CSIRO Parkes Telescope at 10-cm (2.60 - 3.45~GHz). 

Building on these efforts, \citet{wlodarczyk-sroka_extending_2020} refined the constraints on the prevalence of extraterrestrial transmitters by incorporating additional stars within the field of view (defined by the FWHM of the telescope beam) provided by \textit{Gaia} DR2. This furnished a total of 288,315 stars across the 1327 observed pointings: 251,983 at L-band, 172,572 at S-band, 72,030 at 10-cm.

\begin{figure}
\centering
\includegraphics[width=\linewidth]{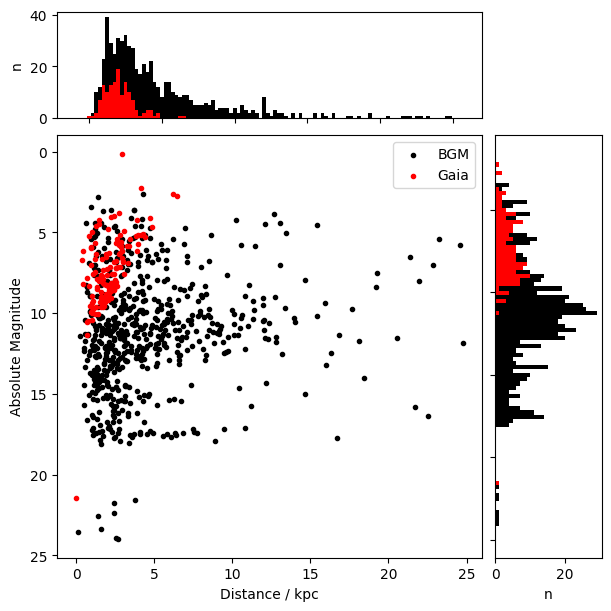}
\caption{The sample of stars within a single pointing (in the direction of l = 276.05 degrees, b = -14.4 degrees) simulated by the BGM (in black) compared to the sample of stars from \textit{Gaia} with accurate parallax measurements (in red). We compare the distribution in absolute magnitudes (right) and distribution of distances (top). The BGM can be used to expand the stellar sample beyond the observational magnitude limitations of \textit{Gaia}, as well as considering the full breadth of the Milky Way.}
\label{fig:example_pointing}
\end{figure}

We have extended the \citet{wlodarczyk-sroka_extending_2020} approach by simulating the stellar bycatch population using the BGM. For 1229 unique pointings from Price's survey, a population of stars was generated without restrictions on the magnitude of stars for distances up to 25~kpc. Two fields located along b $\approx$ 0 degrees towards the Galactic Centre were too dense to simulate using the web portal for the BGM, taking too long to run and so were excluded from our analysis.

Figure~\ref{fig:example_pointing} illustrates the simulated stellar bycatch for a single pointing — towards the target star \textit{Gaia}~DR2~5278042880077383040 — generated using the BGM, and compared against stars observed by \textit{Gaia}. 

Our simulated stellar sample, derived from the 1,229 pointings, includes a total of 6,182,364 stars. This represents a substantial increase compared to the 288,315 stars identified using \textit{Gaia} in the study by \citet{wlodarczyk-sroka_extending_2020}. We utilise the simulated BGM sample to challenge the limitations of using observational catalogues to estimate the stellar bycatch and prevalence of technosignatures.

\subsection{Extending the Stellar Bycatch}

For each simulated star, we calculate the minimum detectable equivalent isotropic radiated power (EIRP$_{\mathrm{min}}$) based on the distance to the star and position with respect to the telescope's field of view. We restrict the field of view to the FWHM of the telescope beam, represented by a Gaussian response. 

The EIRP$_{\mathrm{min}}$ values reported here correspond to the maximum sensitivity achieved when searching for signals with near-zero drift rates. For drift rates exceeding $\pm 0.15$~Hz\,s$^{-1}$, narrowband signals can become smeared across multiple frequency channels—depending on the survey's spectral resolution and integration time—leading to a reduction in sensitivity to such signals \citep{margot_search_2021}. Neither the original analysis by \citet{price_breakthrough_2020} nor the subsequent reanalysis by \citet{wlodarczyk-sroka_extending_2020} account for this effect. To enable a direct comparison with these earlier studies, we have similarly not applied a correction for Doppler smearing in the EIRP$_{\mathrm{min}}$ or CWTFM values presented here. However, we do note that future surveys should correct their results for this effect, as detailed in \citet{margot_search_2021}.

Table \ref{tab:eirp} presents the number of stars simulated by the BGM and observed by \textit{Gaia} for increasing shells of EIRP$_{\mathrm{min}}$, as well as the \emph{Continuous Waveform Transmitter Figure of Merit} (CWTFM). 
CWFTM is a widely used figure of merit in evaluating the effectiveness of radio SETI surveys, and depends on the EIRP$_{\mathrm{min}}$, the fractional bandwidth ($\nu_{rel} = \Delta\nu / \nu_0$), and the total number of stars observed ($N_*$). Following \citet{enriquez_breakthrough_2017} and \citet{price_breakthrough_2020}, the CWTFM can be expressed as:

\begin{equation}
\mathrm{CWTFM} = \zeta_{A0} \frac{\mathrm{EIRP}_{\mathrm{min}}}{N_* \cdot \nu_{rel}} %(\Delta\nu / \nu_0)
\end{equation}

\noindent
Here, $\zeta_{A0}$ is a normalisation constant defined such that $\mathrm{CWTFM} = 1$ corresponds to a benchmark survey: sensitivity to an Arecibo-class transmitter over 1000 stars and a fractional bandwidth of 0.5. 

\begin{table}
\setlength{\arrayrulewidth}{0.2mm}
\renewcommand{\arraystretch}{1}
\centering
\caption{For increasing shells of EIRP$_{\mathrm{min}}$, the stellar bycatch population simulated by the BGM ($N_B$) and observed by Gaia ($N_G$) are compared, and the figure of merit, CWTFM, is estimated for each BGM population. The upper distance limits (d) have been calculated for a star within the centre of the beam (no offset correction applied). The stellar bycatch observed by Gaia has been subject to cuts based on fractional parallax uncertainty and restricted to distances < 10~kpc (see \citet{wlodarczyk-sroka_extending_2020}). The BGM increases the stellar bycatch population for EIRP$_{\mathrm{min}} > 10^{15}$~W.}

\begin{tabular}{c|c|c|c|c|c} 
  & EIRP$_{\mathrm{min}}$ (W) & d (pc) & N$_{B}$ & N$_{G}$ & CWTFM\\ [0.5ex] 
 \hline
 \hline 
 GBT & $10^{14}$ & 290 & 1828 & 2599 & 3.56 \\ [1ex] 
 & $10^{15}$ & 930 & 36071 & 25192 & 1.80 \\ [1ex]
 & $10^{16}$ & 2940 & 367616 & 147976 & 1.77 \\ [1ex]
 & $10^{17}$ & 9310 & 3091788 & 246245 & 2.10 \\ [1ex]
 & $10^{18}$ & 25000 & 5115832 & 246492 & 12.71 \\ [1ex]
 %& $10^{19}$ & 93000 & 5121024 & - & 126.97 \\ [1ex]
\hline
 Parkes & $10^{14}$ & 160 & 98 & 82 & 181.57 \\ [1ex] 
 & $10^{15}$ & 500 & 2019 & 695 & 88.13 \\ [1ex]
 & $10^{16}$ & 1600 & 26613 & 5225 & 66.86 \\ [1ex]
 & $10^{17}$ & 5050 & 267747 & 30434 & 66.46 \\ [1ex]
 & $10^{18}$ & 16000 & 1178528 & 40592 & 150.99 \\ [1ex]
 %& $10^{19}$ & 50500 & 1264841 & - & 1406.83 \\ [1ex]
\end{tabular}
\label{tab:eirp}
\end{table}

The application of the BGM improves the stellar bycatch population - and in turn, the CWFTM - for EIRP$_{\mathrm{min}} \geq 10^{15}$~W. By leveraging the BGM, we are able to account for a much broader stellar population, thereby extending the searchable parameter space to include potential high-power technosignatures originating from across the Milky Way.

\begin{figure}
\centering
\includegraphics[width=0.95\linewidth]{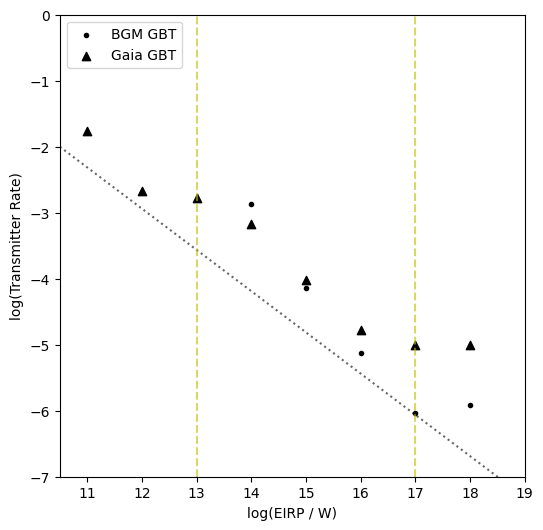}
\caption{For GBT L-band and S-band observations, we plot a comparison of the EIRP$_{\mathrm{min}}$ and transmitter rate determined from the sample of stars up to 25~kpc using the BGM and up to 10kpc from \textit{Gaia} \citep{wlodarczyk-sroka_extending_2020}, for shells of increasing EIRP$_{\mathrm{min}}$. We plot the dashed vertical lines as reference to a transmitter with EIRP$_{\mathrm{min}}$ equivalent to Arecibo ($10^{13}$~W), and a transmitter with equivalent power to a Kardashev Type I civilisation ($10^{17}$~W). This work to extend the stellar bycatch using the BGM challenges the current limits in survey sensitivity and scope, represented in a grey dashed line, referred to as "terra incognito" \citep{garrett_constraints_2023}}
\label{fig:eirp_tr}
\end{figure}

Considering how increasing the stellar bycatch improves the EIRP$_{\mathrm{min}}$ - Transmitter Rate parameter space, we plot figure \ref{fig:eirp_tr} for GBT's EIRP$_{\mathrm{min}}$ shells. Beyond 100~pc, application of the BGM improves the transmitter rate against increasing shells of EIRP$_{\mathrm{min}}$; simulating the stellar bycatch population beyond magnitude thresholds challenges the limits of 'terra incognito' \citep{garrett_constraints_2023} and current searched parameter space, in terms of volume of sky searched and the sensitivity depth.

The BGM provides a powerful framework for estimating the underlying stellar population and for simulating the total stellar bycatch missed by \textit{Gaia}. However, we note that within $\sim 100$~pc the BGM simulates fewer stars along the 1229 pointings than that  detected by \textit{Gaia} (see Figure~\ref{fig:eirp_tr}). Consequently, we underestimate the number of stars probed by the SETI surveys considered here for transmissions with EIRP$_{\mathrm{min}} \leq 10^{14}$~W. From these results, it looks as though the BGM underestimates the number of stars simulated on these very limited spatial scales by a factor of $\sim 30$\%. We also note that, when comparing against \citet{wlodarczyk-sroka_extending_2020}, the population of stars with EIRP$_{\mathrm{min}} \leq 10^{14}$~W contains the targets from the Enriquez/Price survey.

Beyond this range, however, the BGM provides a significantly more complete representation of the underlying stellar population, including far-field stellar objects that might otherwise be filtered out of observational catalogues due to significant astrometric uncertainties. It is therefore the preferred tool for extending bycatch estimates to larger Galactic volumes.

\subsection{Diversity beyond Observational Magnitude Limits}

\begin{figure}
\centering
\includegraphics[width=0.95\linewidth]{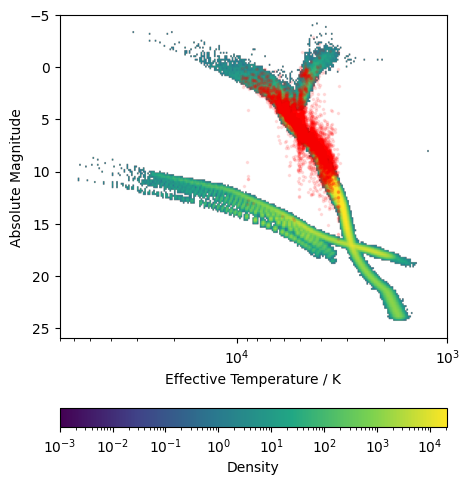}
\caption{Hertzsprung–Russell diagram comparing the BGM-simulated stellar population (background) with the subset of \textit{Gaia} sources (superimposed in red) for which reliable effective temperatures and absolute magnitudes are available. The \textit{Gaia} sample is restricted to within 1~kpc due to parallax accuracy, and most white dwarfs lack reliable temperature estimates. Utilising the BGM enables consideration of the complete breadth of stellar types captured within the bycatch population, including bright Main Sequence stars and minimal confusion between faint Main Sequence and white dwarfs.}
\label{fig:hr_diag}
\end{figure}

Galactic modelling also allows us to explore the true statistical diversity of stellar types sampled in the Enriquez/Price surveys. By plotting effective temperature against absolute magnitude in a Hertzsprung–Russell (HR) diagram (Figure~\ref{fig:hr_diag}), we can examine the physical properties of the stellar populations simulated by the BGM. The simulation extends coverage across both extremes of the Main Sequence and reveals a rich population of white dwarfs. 
We can avoid photometric confusion in spectral classification with \textit{Gaia} \citep{arenou_gaia_2018}, placing clear cuts between faint Main Sequence stars and white dwarfs using the simulated bycatch. 

\begin{figure}
\centering
\includegraphics[width=0.95\linewidth]{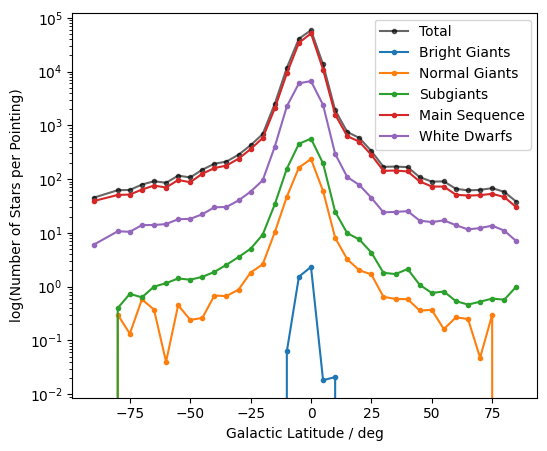}
\caption{The distribution of stellar spectral types as a function of galactic latitude for BGM simulations of the 1229 pointing positions. While the overall diversity peaks near the Galactic Plane as expected, the relative proportions of stellar types remain broadly stable across all latitudes.}
\label{fig:spec_types}
\end{figure}

\begin{figure}
\centering
\begin{subfigure}{0.95\linewidth}
\includegraphics[width=\linewidth]{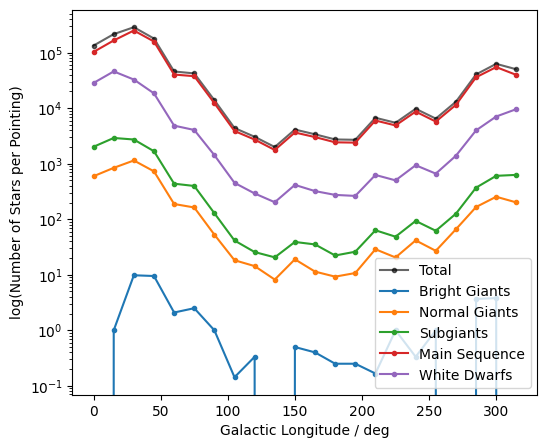}
\label{fig:latzero}
\end{subfigure}
\begin{subfigure}{0.95\linewidth}
\includegraphics[width=\linewidth]{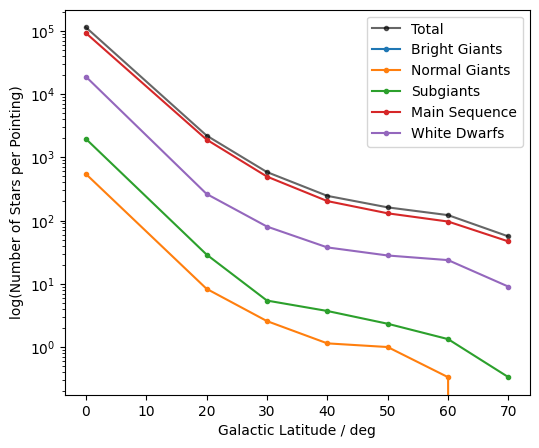}
\label{fig:longzero}
\end{subfigure}
\caption{The distribution of stellar spectral types in relation to the Galactic Plane. (Upper) The number of stars per pointing per spectral type for $b < |5^\circ|$. (Lower) The number of stars per pointing per spectral type for $l < |5^\circ|$. Both figures highlight the opportunistic bycatch towards the Galactic Centre for SETI endeavours.}
\label{fig:gal_centre_types}
\end{figure}

To assess how this spectral diversity varies spatially, we analyse the distribution of stellar types as a function of Galactic latitude in figure~\ref{fig:spec_types}. As expected, the greatest diversity and density of stellar types is found in the Galactic Plane. However, the overall ratio of spectral types remains broadly consistent across latitudes. Restricting the sample of pointings to galactic latitudes within $\pm 5^\circ$ and galactic longitudes within $\pm 5^\circ$ of the Galactic Centre in Figure~\ref{fig:gal_centre_types}, we provide further support, on grounds of stellar density and diversity, for continuing technosignature searches toward the inner Galaxy, particularly the Galactic Centre \citep{gajjar_breakthrough_2021}. While we note that this region may not be the most conducive environment for the development and sustainability of life, it remains probabilistically opportunistic for future SETI endeavours.  

This modelling also highlights the diverse range of stellar types that are inadvertently included in single-dish SETI observations. Even when surveys are designed to focus on specific stellar classes, such as nearby G-type Main Sequence stars, they inevitably include a broader array of stellar types as a by-product. As a result, single-dish SETI surveys, regardless of their specific targets, implicitly impose constraints on a wide variety of stellar types, including those not specifically aimed at by the principal investigator. Given the Galaxy's environmental diversity and the limited understanding of the conditions under which technological civilizations might arise, it is important to recognize these implications and consider the valuable constraints that SETI surveys inherently impose on a wide spectrum of stellar hosts. Using the BGM to estimate the stellar bycatch in SETI surveys allows us to explicitly assess and constrain the prevalence of extraterrestrial transmitters across a diverse range of stellar types.

\subsection{Estimating the Prevalence of ETI}

A key goal of SETI surveys is to assess what their null detection results imply about the prevalence of extraterrestrial intelligence (ETI) within the Galaxy. When considering the detection of a technosignature as an independent event, and given the large number of stars encompassed within each pointing, the probability distribution of transmitter detections can be modelled as a Poisson process \citep{choza_breakthrough_2024}. 

For a Poisson-distributed random variable, the probability of observing $k$ events, given an expected mean rate $\lambda$, is given by
\[
P(k | \lambda) = \frac{\lambda^{k} e^{-\lambda}}{k!}.
\]
In the case of a null detection ($k = 0$), this simplifies to
\[
P(0 | \lambda) = e^{-\lambda}.
\]
To set an upper bound on $\lambda$ at a given confidence level $C$, we require
\[
P(k \leq 0 | \lambda_{\mathrm{upper}}) = e^{-\lambda_{\mathrm{upper}}} = 1 - C.
\]
Rearranging gives
\[
\lambda_{\mathrm{upper}} = -\ln(1 - C).
\]
For the commonly used 95\% confidence interval, $C = 0.95$, and thus
\[
\lambda_{\mathrm{upper}} \approx -\ln(0.05) \approx 3.
\]
This is the well-known ``Rule of Three’’ which allows us to estimate the upper bound of the mean detection rate \citep{gehrels_confidence_1986} as 
\[
\lambda \approx \frac{3}{n},
\]
where $n$ is the number of independent target stars surveyed. (Please see \citep{margot_search_2023} for alternative binomial derivation)

\begin{figure}
\centering
\includegraphics[width=0.95\linewidth]{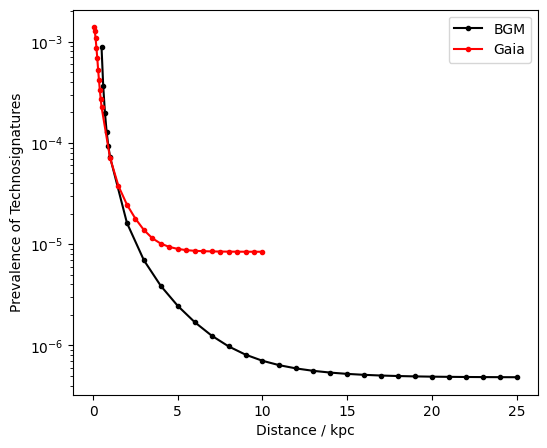}
\caption{Estimates of the prevalence of transmitters within the Milky Way, based on the upper bound of the 95\% confidence interval, assuming Poisson statistics for the detection of a technosignature and a continuous high duty cycle transmitter. By simulating the stellar bycatch, we vastly improve the stellar bycatch population and restrict the prevalence estimated using Gaia beyond 2.5~kpc.}
\label{fig:prev}
\end{figure}

Applying this to our dataset, we find no evidence for continuous, high duty cycle transmitters within the stellar bycatch of the 1229 unique pointings analysed. For distances up to 2.5~kpc, we simulate $n = 301,472 \pm 549$ stars, leading to an estimated prevalence of such transmitters of
\[
f_{\mathrm{tx}} \leq \frac{3}{n} \approx (0.000995 \pm 0.000002)\%,
\]
for transmitters with $\mathrm{EIRP}_{\mathrm{min}} \gtrsim 5 \times 10^{16}\,\mathrm{W}$. 

By extending this analysis to concentric shells out to 25~kpc in figure \ref{fig:prev}, we estimate the prevalence of high duty cycle extraterrestrial transmitters at increasing Galactic radii. Importantly, our simulations of the stellar bycatch population provide a significantly more constrained estimate of transmitter prevalence beyond 2.5~kpc than previous \textit{Gaia}-based approaches, while accounting for a broad range of stellar spectral types (see Appendix~\ref{appendix}).

\section{Application to Future SETI Surveys}

For the ease of application to future SETI surveys, we have developed a simple calculator to run simulations via the BGM's web service \footnote{The calculator code can be found at \url{https://github.com/louisam11/SETI-Stellar-Bycatch-Simulator}}. The user runs the simulation, via the BGM web portal \footnote{Users are required to create an account to run BGM simulations, to enable access to the simulations through \url{https://model.obs-besancon.fr/} as well as to receive an email notification when the simulation is complete} by providing a pointing direction (in Right Ascension and Declination) and a telescope field of view (we chose to use the FWHM for the field of view but other values are possible) for simulations that generate $\leq$ 2~gigabytes. There is a time delay between job submission and completion, depending on the line of sight for the pointing and population size to be synthesised. 

\begin{figure}
\centering
\includegraphics[width=0.95\linewidth]{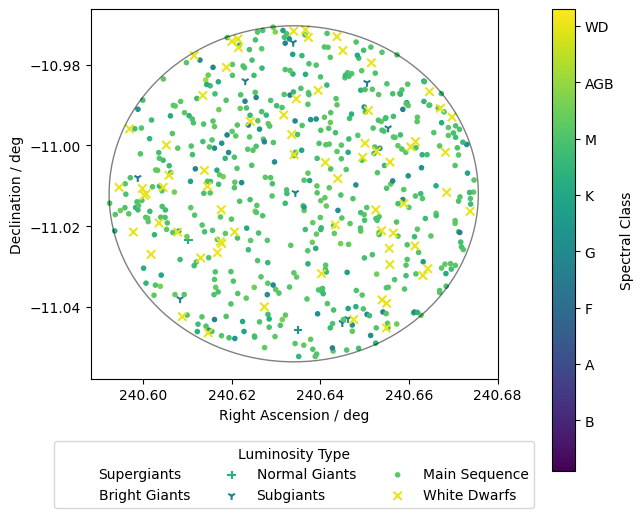}
\caption{A demonstration of a typical field of view plot, produced by the calculator for a stellar bycatch sample captured within a 5" field of view, in the direction of (l = 0$^\circ$, b = 30$^\circ$). The diversity in spectral classes (colour-coded) and luminosity types (differentiated by marker) has been highlighted. Note no supergiants or bright giants were found within this particular pointing.}
\label{fig:fov}
\end{figure}

The BGM simulates the stellar population for a rectangular solid angle (in square degrees), and our calculator then filters this down to only include stars with the radial field of view requested by the user. The BGM produces two output files: one containing the main stellar characteristics and another header file that records the parameters used for the simulation. The calculator extracts key stellar properties from the results (in ASCII format), including position, magnitude and colour, age, mass, effective temperature, and spectral classification. An example of this is shown in table \ref{tab:example}. 
The EIRP$_\mathrm{{min}}$ for each star is estimated given the inputted telescope's System Equivalent Flux Density (SEFD), spectral resolution, signal-to-noise threshold and integration time. With a dechirping efficiency specified by the user, we consider the potential doppler smearing \citep{margot_search_2021} as a result of searching at high drift rates and correct the EIRP$_\mathrm{{min}}$ for the maximum sensitivity loss. The calculator will generate a plot of the stellar bycatch population captured within the field of view (see  figure \ref{fig:fov} for an example), and highlights the diversity of spectral classes and luminosity types captured per pointing.

\begin{table*}
\setlength{\arrayrulewidth}{0.2mm}
\renewcommand{\arraystretch}{1}
\centering
\caption{An example output from the BGM, containing the first three stars simulated for pointing direction (l = 0$^\circ$, b = 30$^\circ$) and a field of view of 5". Other user input values assume the use of the GBT S-band receiver for 5 minute observations, with dechirping efficiency of 16.5\% and SNR of 10.}
\begin{tabular}{c|c|c|c|c|c|c|c|c|c|c|c|c|c|c} 
RA & Dec & Dist & M$_V$ & m$_G$ & G-V & Age & Mass & T$_{\mathrm{eff}}$ &  Spectral & Luminosity & EIRP$_{\mathrm{min}}$ & EIRP$_{\mathrm{min, smear}}$ \\ [0.5ex] 
(deg) & (deg)  & (pc) &  &  &  & ($10^{9}$ yrs) & (M$_\odot$) & (K) & Class  & Class & (W) & (W) \\ [0.5ex]
 \hline
240.6209 & -10.9977 & 314.1 & 10.44 & 18.84 & 0.830 & 3.958 & 0.499 & 3516 & M2 & Main Sequence & 2.318 $\times 10^{14}$ & 1.405 $\times 10^{15}$\\ [1ex] 
240.6317 & -10.9924 & 347.1 & 17.70 & 25.54 & 0.037 & 9.961 & 0.577 & 2324 & WD DB & White Dwarfs & 2.830 $\times 10^{14}$ & 1.715 $\times 10^{15}$\\ [1ex] 
240.6674 & -10.9944 & 362.0 & 6.42 & 14.58 & 0.242 & 1.434 & 0.751 & 4687 & K2 & Main Sequence & 3.088 $\times 10^{14}$ & 1.872 $\times 10^{15}$\\ [1ex] 
\vdots &  &  & \vdots &  &  &  & \vdots &  &  &  & \vdots \\ [1ex]
\end{tabular}
\label{tab:example}
\end{table*}

Generating a population of stars without limitations on magnitude and within 25~kpc, the calculator sub-divides the bycatch population depending on their EIRP$_{\mathrm{min}}$. The EIRP$_{\mathrm{min}}$ has been corrected for the telescope's response, assuming a 2D Gaussian profile across the FWHM of the beamwidth for a single-dish telescope. The calculator generates a table for increasing EIRP$_{\mathrm{min}}$ shells of the estimated transmitter rate, based on the number of stars and the fractional bandwidth, and plots the bycatch sample in EIRP$_{\mathrm{min}}$-Transmitter Rate parameter space. 

The calculator then sub-divides the bycatch population by luminosity class and spectral type, and estimates the prevalence per stellar class for shells of increasing distance, up to 10~kpc (similar to table \ref{tab:prev}). The breadth of diversity captured within the sample can be visually represented through the HR diagram for the stellar bycatch population simulated. Finally, the calculator estimates the prevalence of technosignatures based on the upper bound of the 95\% confidence interval, assuming Poisson statistics for the detection of a technosignature and a continuous high duty cycle transmitter (generating a plot similar to figure \ref{fig:prev}).

\section{Conclusion}

This paper builds on the approach of \citet{wlodarczyk-sroka_extending_2020} by emphasising the importance of including the stellar bycatch for targeted radio SETI observations.
%- the incidental stellar population, located in the periphery and background of targeted radio SETI observations. 
By simulating stars within the field of view using the BGM, we can overcome the observational limitations of \textit{Gaia} and improve upon statistical estimates of the prevalence of extraterrestrial transmitters across a broader stellar population, including lines of sight susceptible to confusion and crowding.

We demonstrate the use of the BGM approach through re-analysis of the Enriquez and Price surveys, simulating a sample of 6,182,364 stars across 1,229 unique pointings. The BGM significantly increases the number of stars considered above EIRP$_{\mathrm{min}} > 10^{15}$~W. We place limits on the prevalence of high duty cycle transmitters within 2.5~kpc, suggesting $\leq (0.000995 \pm 0.000002)$\% of stellar systems contain such a transmitter (for EIRP$_{\mathrm{min}} \gtrsim 5 \times 10^{16}$~W).
This approach shows that single-dish surveys can place far stronger constraints on the prevalence of extraterrestrial transmitters across a wide diversity of stellar types. By incorporating this diversity through BGM simulations, we find that SETI surveys are, in fact, less biased by anthropocentric assumptions than often suggested. 

We provide a calculator that enables users to identify the simulated stellar bycatch population for future single-dish SETI surveys. In addition, this approach is applicable to incoherent beamforming surveys (such as COSMIC on the VLA \citep{tremblay_cosmics_2025} or BLUSE on MeerKAT \citep{czech_breakthrough_2021}) and interferometric SETI (iSETI) (such as \citet{wandia_interferometric_2023} and \citet{mason_conducting_2025}). Incorporating more realistic constraints on the prevalence of extraterrestrial life promotes best practice in quantifying the true extent of SETI’s uncharted parameter space and will help inform the design and interpretation of future surveys.

\section*{Acknowledgements}
We would like to thank Dr Eamonn Kerins for invaluable discussion, as well as assistance from the BGM team (in particular Dr Céline Reylé) with the application of the BGM. The calculator utilises code written by Raphael Meloir to launch simulations for the BGM web service and Gravpot web service. 

\appendix \section{Prevalence Estimates by Spectral Type} \label{appendix}
We consider the spectral diversity for shells of increasing distance (up to 10~kpc) for 1229 pointings. The prevalence is estimated for spectral shells where $N_* \geq 50$, given that the uncertainty in number of stars is $\sqrt{N_*}$. Through the use of BGM simulations, we consider a broader range of potential sources for high duty cycle technosignatures when estimating the bycatch population.

\begin{table*}
\setlength{\arrayrulewidth}{0.2mm}
\renewcommand{\arraystretch}{1}
\centering
\caption{We estimate the prevalence of ETI transmitters, $p$, for shells of individual spectral classifications and distances $\leq$ 10~kpc, where $N_* \geq 50$.}
\begin{tabular}{cc|cc|cc|cc|cc|} 
 & & \multicolumn{2}{c}{500~pc} & \multicolumn{2}{c}{1~kpc} & \multicolumn{2}{c}{5~kpc} & \multicolumn{2}{c}{10~kpc} \\ [0.5ex] 
Type & Class & N$_*$ & p (\%) & N$_*$ & p (\%) & N$_*$ & p (\%) & N$_*$ & p (\%) \\ [0.5ex] 
\hline

Bright Giants & G & 0 & - & 0 & - & 16 & - & 60 & < 5.0 $\pm$ 0.6 \\ [1ex]
 & K	& 0 & - & 1 & - & 25 & - & 57 & < 5.3 $\pm$ 0.7 \\ [1ex]

Normal Giants & A & 0 & - & 6 & - & 124 & < 2.4 $\pm$ 0.2 & 403 & < 0.74 $\pm$ 0.04 \\ [1ex]
 & F & 0 & -	& 10 & - & 216 & < 1.4 $\pm$ 0.1 & 669 & < 0.45 $\pm$ 0.01 \\ [1ex]
 & G & 5 & - & 67 & < 4.5 $\pm$ 0.5 & 2035 & < 0.147 $\pm$ 0.003 & 8308 & < 0.036 $\pm$ 0.001 \\ [1ex]
 & K	& 3 & - & 80 & < 3.8 $\pm$ 0.4 & 2362 & < 0.127 $\pm$ 0.003 & 7561 & < 0.040 $\pm$ 0.001 \\ [1ex]
 & M	& 0 & - & 6 & - & 88 & < 3.4 $\pm$ 0.4 & 308 & < 0.97 $\pm$ 0.06 \\ [1ex]

Subgiants & B & 0 & - & 0 & - & 39 & - & 106 & < 2.8 $\pm$ 0.3 \\ [1ex]
 & A	& 0 & - & 10 & - & 242 & < 1.2 $\pm$ 0.1 & 727 & < 0.41 $\pm$ 0.02 \\ [1ex]
 & F & 23 & - & 305 & < 0.98 $\pm$ 0.06 & 10169 & < 0.030 $\pm$ 0.0003 & 41058 & < 0.007 $\pm$ 4 $\times 10^{-5}$ \\ [1ex]
 & G & 6 & - & 82 & < 3.7 $\pm$ 0.4 & 2291 & < 0.131 $\pm$ 0.003 & 7101 & < 0.042 $\pm$ 0.0005 \\ [1ex]

Main Sequence & B & 0 & - & 3 & - & 91 & < 3.3 $\pm$ 0.3 & 315 & < 0.95 $\pm$ 0.05 \\ [1ex]
 & A & 1 & - & 25 & - & 657 & < 0.46 $\pm$ 0.02 & 2044 & < 0.147 $\pm$ 0.003 \\ [1ex]
 & F & 69 & < 4.3 $\pm$ 0.5 & 585 & < 0.51 $\pm$ 0.02 & 19179 & < 0.016 $\pm$ 0.0001 & 75744 & < 0.004 $\pm$ 1 $\times 10^{-5}$ \\ [1ex]
 & G & 257 & < 1.2 $\pm$ 0.1 & 1768 & < 0.170 $\pm$ 0.004 & 54715 & < 0.005 $\pm$ 2 $\times 10^{-5}$ & 207808 & < 0.001 $\pm$ 3 $\times 10^{-6}$\\ [1ex] 
 
 & K & 611 & < 0.49 $\pm$ 0.02 & 4258 & < 0.070 $\pm$ 0.001 & 125692 & < 0.002 $\pm$ 7 $\times 10^{-6}$ & 501400 & < 0.0005 $\pm$ 8 $\times 10^{-7}$ \\ [1ex]
 & M & 6059 & < 0.050 $\pm$ 0.001 & 32328 & < 0.009 $\pm$ 5 $\times 10^{-5}$ & 820425 & < 0.0004 $\pm$ 4 $\times 10^{-7}$ & 2696160 & < 0.0001 $\pm$ 7 $\times 10^{-8}$ \\ [1ex]
 & AGB & 251 & < 1.2 $\pm$ 0.1 & 1316 & < 0.23 $\pm$ 0.01 & 28453 & < 0.011 $\pm$ 6 $\times 10^{-5}$ & 81198 & < 0.004 $\pm$ 1 $\times 10^{-5}$ \\ [1ex]

White Dwarfs & DA & 310 & < 0.97 $\pm$ 0.06 & 2302 & < 0.130 $\pm$ 0.003 & 92980 & < 0.003 $\pm$ 1 $\times 10^{-5}$ & 442050 & < 0.0007 $\pm$ 1 $\times 10^{-6}$ \\ [1ex]
 & DB & 236 & < 1.3 $\pm$ 0.1 & 1381 & < 0.22 $\pm$ 0.01 & 32809 & < 0.009 $\pm$ 5 $\times 10^{-5}$ & 92144 & < 0.003 $\pm$ 1 $\times 10^{-5}$ \\ [1ex]
 & DC & 235 & < 1.3 $\pm$ 0.1 & 1412 & < 0.21 $\pm$ 0.01 & 32711 & < 0.009 $\pm$ 5 $\times 10^{-5}$ & 91644 & < 0.003 $\pm$ 1 $\times 10^{-5}$ \\ [1ex]

\end{tabular}
\label{tab:prev}
\end{table*}

\section*{Data Availability}
This work presents results from the European Space Agency (ESA) space mission \textit{Gaia}. \textit{Gaia} data are being processed by the \textit{Gaia} Data Processing and Analysis Consortium (DPAC). Funding for the DPAC is provided by national institutions, in particular the institutions participating in the \textit{Gaia} MultiLateral Agreement (MLA). The \textit{Gaia} mission website is https://www.cosmos.esa.int/gaia. The \textit{Gaia} archive website is https://archives.esac.esa.int/gaia. This research has made use of the VizieR catalogue access tool, CDS, Strasbourg, France \citep{ochsenbein_vizier_2000}. This work made use of \textsc{astropy}, a community-developed core \textsc{python} package and an ecosystem of tools and resources for astronomy (Astropy Collaboration \citeyear{astropy:2013}, \citeyear{astropy:2018}, \citeyear{astropy:2022}).



\bibliographystyle{mnras}
\bibliography{References.bib} % if your bibtex file is called example.bib

%%%%%%%%%%%%%%%%%%%%%%%%%%%%%%%%%%%%%%%%%%%%%%%%%%

%%%%%%%%%%%%%%%%% APPENDICES %%%%%%%%%%%%%%%%%%%%%

%%%%%%%%%%%%%%%%%%%%%%%%%%%%%%%%%%%%%%%%%%%%%%%%%%

% Don't change these lines
\bsp	% typesetting comment
\label{lastpage}
\end{document}